\documentclass{article}

\usepackage[nonatbib, final]{neurips_2021_ml4ps}
\usepackage{float}
\usepackage{chngpage}
\usepackage{xcolor}

\usepackage[utf8]{inputenc} 
\usepackage[T1]{fontenc}    
\usepackage{hyperref}       
\hypersetup{
    colorlinks=true,
    linkcolor=black,
    filecolor=magenta,      
    urlcolor=blue,
    citecolor=blue,
}

\usepackage{url}            
\usepackage{booktabs}       
\usepackage{topcapt}        
\usepackage{amsfonts}       
\usepackage{nicefrac}       
\usepackage{microtype}      
\usepackage{lipsum}
\usepackage{graphicx}
\usepackage{amsmath}
\usepackage{multirow}
\usepackage{xspace}
\usepackage{cite}
\usepackage{listings}

\usepackage{color}
\definecolor{deepblue}{rgb}{0,0,0.5}
\definecolor{deepred}{rgb}{0.6,0,0}
\definecolor{deepgreen}{rgb}{0,0.5,0}

\usepackage{subcaption}
\usepackage{array}
\newcolumntype{-}{>{\global\let\currentrowstyle\relax}}
\newcolumntype{^}{>{\currentrowstyle}}

\DeclareFixedFont{\ttb}{T1}{txtt}{bx}{n}{10} 
\DeclareFixedFont{\ttm}{T1}{txtt}{m}{n}{10}  

\newcommand\pythonstyle{\lstset{
language=Python,
basicstyle=\ttm,
otherkeywords={def},             
keywordstyle=\ttb\color{deepblue},
emph={MyClass,__init__},          
emphstyle=\ttb\color{deepred},    
stringstyle=\color{deepgreen},
frame=tb,                         
showstringspaces=false            %
}}

\lstnewenvironment{python}[1][]
{
\pythonstyle
\lstset{#1}
}
{}

\newcommand\pythoninline[1]{{\pythonstyle\lstinline!#1!}}

\renewcommand{\vec}[1]{\mathbf{#1}}

\title{Explaining machine-learned particle-flow reconstruction}

\newcommand{\pt}{\ensuremath{p_{\mathrm{T}}}\xspace}

\newcommand{\TeV}{\ensuremath{\,\text{Te\hspace{-.08em}V}}\xspace}

\newcommand{\PYTHIA} {{\textsc{pythia}}\xspace}

\newcommand{\ttbar}{\ensuremath{\mathrm{t}\overline{\mathrm{t}}}\xspace}

\newcommand{\DELPHES} {{\textsc{delphes}}\xspace}

\begin{document}

\author{
  Farouk Mokhtar, Raghav Kansal, Daniel Diaz, Javier Duarte\\
  University of California San Diego \\
  La Jolla, CA 92093, USA \\
  \And
   Joosep Pata\\
   National Institute of Chemical Physics and Biophysics (NICPB)\\ 
   R\"{a}vala pst 10, 10143 Tallinn, Estonia
   \And
  Maurizio Pierini\\
  European Organization for Nuclear Research (CERN) \\
  CH-1211 Geneva 23, Switzerland
  \And
   Jean-Roch Vlimant\\
   California Institute of Technology \\
   Pasadena, CA 91125, USA
}

\maketitle

\begin{abstract}
The particle-flow (PF) algorithm is used in general-purpose particle detectors to reconstruct a comprehensive particle-level view of the collision by combining information from different subdetectors. 
A graph neural network (GNN) model, known as the machine-learned particle-flow (MLPF) algorithm, has been developed to substitute the rule-based PF algorithm. 
However, understanding the model's decision making is not straightforward, especially given the complexity of the set-to-set prediction task, dynamic graph building, and message-passing steps.
In this paper, we adapt the layerwise-relevance propagation technique for GNNs and apply it to the MLPF algorithm to gauge the relevant nodes and features for its predictions.
Through this process, we gain insight into the model's decision-making.
\end{abstract}


\section{Introduction}
Machine learning (ML) has been used in many facets of Large Hadron Collider (LHC) computing and physics analyses.
It can provide a faster and more accurate paradigm for designing reconstruction algorithms, compared to hand-coded, rule-based, imperative algorithms. 
Graph neural networks (GNNs), which are well-suited to non-Euclidean structured data such as detector data, have been shown to be performant for high energy physics (HEP) tasks~\cite{Shlomi:2020gdn,Duarte:2020ngm} ranging from charged particle tracking~\cite{Farrell:2018cjr,Ju:2020xty,Amrouche:2019wmx,Amrouche:2019yxv,Choma:2020cry,Dezoort:2021kfk}, to jet classification~\cite{Moreno:2019bmu,Moreno:2019neq,Qu:2019gqs,Mikuni:2020wpr}, clustering~\cite{Qasim:2019otl,Kieseler:2020wcq}, and simulation~\cite{kansal2021particle}.

An important core software algorithm at the LHC is the particle-flow (PF) algorithm~\cite{Behrend:1982gk,Buskulic:1994wz,H1:2020zpd,Breitweg:1997aa,Breitweg:1998gc,Abreu:1995uz,Bocci:2001zx,Connolly:2003gb,Abulencia:2007iy,Abazov:2008ff,Sirunyan:2017ulk,Aaboud:2017aca}, which takes heterogeneous detector information, namely charged particle tracks~\cite{Strandlie:2010zz,Chatrchyan:2014fea,Aaboud:2017all} and calorimeter energy clusters~\cite{fabjan:1982,Wigmans:1990ey,Fabjan:2003aq}, and returns a list of final-state particles known as PF candidates. 

In a previous effort, the machine-learned particle-flow (MLPF)~\cite{Pata:2021oez} algorithm was developed as an end-to-end trainable substitute based on parallelizable, computationally efficient, and scalable GNNs. 
Those results show that the MLPF algorithm gives comparable performance to a rule-based PF algorithm.
In this paper, we expand on this by applying layerwise-relevance propagation (LRP)~\cite{LRP,Montavon2019}, an explainable artificial intelligence (XAI)~\cite{XAI} technique, to MLPF. 

ML models, which operate non-trivially on high-dimensional representations of input features often suffer from a lack of interpretability.
Understanding the decision-making behind the PF-candidate class and feature predictions is extremely valuable in order to increase confidence, ensure robustness under changing conditions, and enable us to potentially glean new insights about the detector performance or reconstruction not currently utilized by the rule-based PF algorithm.
The term ``black box'' has long been associated with ML algorithms as they operate on, and learn directly from, data in ways that are not obvious to humans. 
In recent years, the interpretability of ML models have received increasing attention by both the academic research community and industry under the umbrella of XAI~\cite{XAI}. 

This paper is organized as follows. 
Section~\ref{sec:related} reviews related work, while \autoref{sec:Modifications} describes the modifications we made to the LRP technique in order to apply it to GNNs, in general, and specifically to the MLPF algorithm.
Section~\ref{sec:Results} describes the dataset and the results in terms of the relevancy scores of different features.
Finally, \autoref{sec:summary} summarizes our findings.

\section{Related Work}
\label{sec:related}

\paragraph{XAI in HEP}
\label{sec:LRPhep}
One application of LRP in HEP is a framework presented to extract and understand decision-making information from a DNN classifier of jet substructure tagging techniques~\cite{Agarwal:2020fpt}. 
The LRP technique was combined with expert augmented variables to not only find the relevant network information, but also rank features, potentially highlighting a reduced set that capture the network performance.
Another application of XAI in HEP is a white box AI~\cite{Lai:2020byl} approach which demonstrated that a generative adversarial network (GAN) was able not only to learn the final distribution of partons in an event but also extract the underlying physics using full event information registered in the detectors.

\paragraph{XAI on Graphs}
\label{sec:LRPgraphs}
One attempt to develop a consistent LRP application on graphs for a graph classification task is the GNN-LRP method~\cite{GNNLRP} which proposes explanations in the form of scored sequences of edges on the input graph.
Another attempt is the graph layerwise-relevance propagation (GLRP)~\cite{GLRP} method which extends the procedure of LRP to make it available for Graph-CNN~\cite{GraphCNN} and test its applicability on a large breast cancer dataset. 
This was done for a graph signal classification task, which is different from the problem at hand.
Another XAI method developed to operate on graphs is the GNNExplainer~\cite{GNNExplainer} which attempts to highlight a compact subgraph structure and a small subset of node features that are most relevant for a GNN's prediction. 
It attempts to maximize the mutual information between a GNN's prediction and distribution of possible subgraph structures. 
We build upon existing LRP methods to adapt them to the GravNet layer in MLPF.

\section{LRP Applied to the MLPF Algorithm}
The MLPF algorithm is based on the GravNet~\cite{Qasim:2019otl} layer, which learns a high-level embedding of the node positions and performs dynamic graph construction by connecting the $k=16$ nearest neighbors in this embedding space. 
It receives as input a set of detector elements $X=\{x_i\}$ per event (up to 5,000 in LHC Run 2 conditions) and attempts to predict the corresponding set of PF candidates $Y=\{y_i\}$. 
The set $X$ can be best represented as a heterogeneous point cloud in feature space due to the different elements---from different subdetectors---and inherently non-Euclidean structure of the detector. 
Detector elements and target PF candidates are specified by
\begin{align}
x_i &= [\mathrm{type}, \pt, \eta, \phi, \eta_\mathrm{out},  \phi_\mathrm{out}, E_\mathrm{ECAL},  E_\mathrm{HCAL}, \mathrm{charge}, \mathrm{is\_gen\_el}, \mathrm{is\_gen\_mu}]\\
y_i &= [\mathrm{PID}, \pt, E, \eta, \phi, \mathrm{charge}]
\end{align}
where type $\in$ \{cluster, track\}, PID $\in$ \{null, charged hadron, neutral hadron, photon, electron, muon\}, $E_\mathrm{ECAL}$ ($E_\mathrm{HCAL}$) is the energy in the electromagnetic (hadron) calorimeter, and $\eta_\mathrm{out}$ and $\phi_\mathrm{out}$ are the extrapolations of $\eta$ (pseudorapidity) and $\phi$ (azimuthal angle) to the tracker edge. 
For input tracks, only the type, $\pt$ (transverse momentum), $\eta$, $\phi$, $\eta_\mathrm{out}$, $\phi_\mathrm{out}$, and charge features are filled. 
Similarly, for input clusters, only the type, $E_\mathrm{ECAL}$, $E_\mathrm{HCAL}$, $\eta$,  $\phi$, entries are filled.
As is done in Ref.~\cite{Pata:2021oez}, the target set $Y$ is zero-padded such that $|Y|=|X|$, and $Y$ is then arranged such that if a target PF candidate can be associated geometrically to a detector element input, it is arranged to be in the same location in the sequence.

The tasks at hand are both the classification of PF candidates and the regression of their kinematics.
The MLPF model is implemented and trained using PyTorch Geometric~\cite{PyTorchGeometric}, which accommodates variable-sized input graphs.

\begin{figure}[ht]
    \centering
    \includegraphics[width=0.71\textwidth]{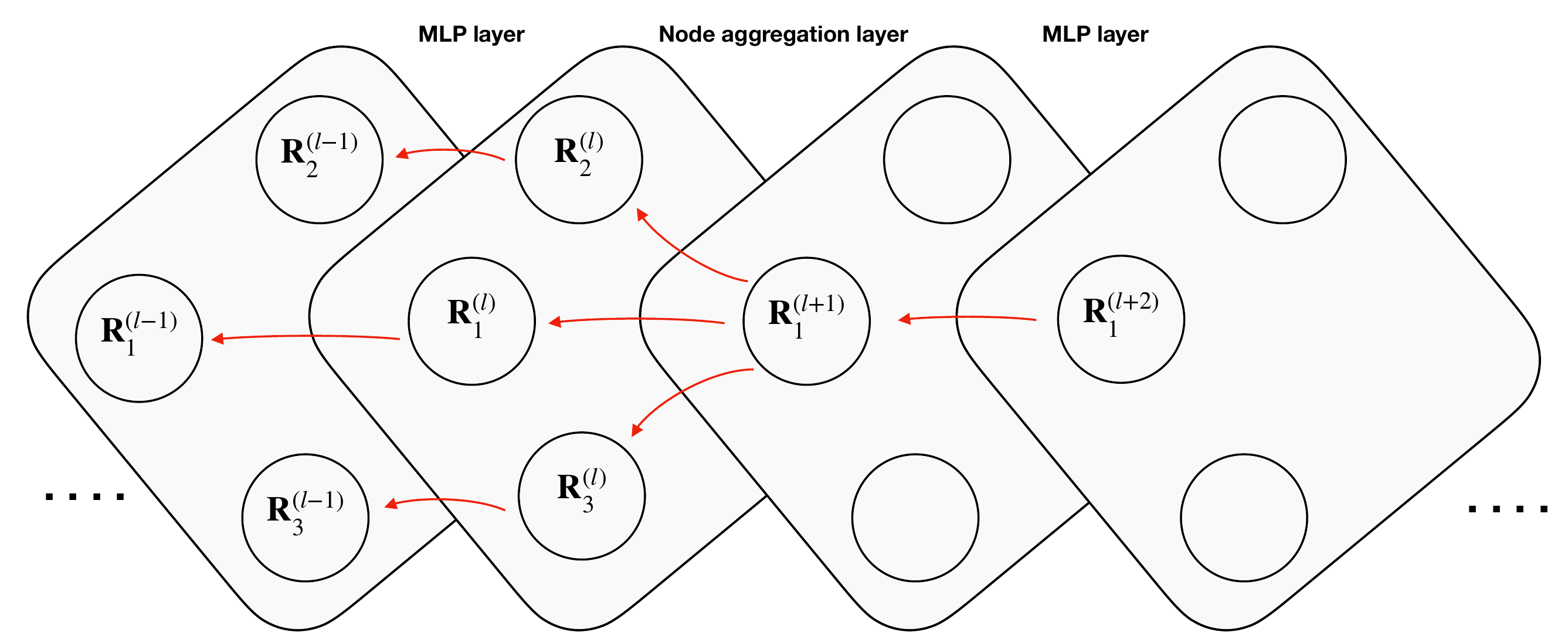}
    \caption{The flow of $R$-scores of node 1 across the different layers in MLPF. 
    For MLP layers, the redistribution of $R$-scores follows the standard LRP rules~\cite{LRP,Montavon2019}. 
    For the aggregation step in the message passing layer, the redistribution follows \autoref{eq:LRP}. 
    We only show three nodes for simplicity.}
    \label{fig:LRP_msg}
\end{figure}

\paragraph{LRP Rules}
\label{sec:Modifications}
MLPF largely consists of multilayer perceptron (MLP) transformations, to which the standard LRP rules~\cite{LRP,Montavon2019} are applicable, with the additional operation of aggregating messages from neighboring nodes in a graph.
In this step, as shown in \autoref{fig:LRP_msg}, we distribute the relevance scores per node post-aggregation across the nodes pre-aggregation according to the following formula:
\begin{equation}\label{eq:LRP}
    \vec{R}^{(l)}_j = \sum_{k}^{}\frac{x_{j}A_{jk}}{\sum_{m}^{}x_{m}A_{mk}}\vec{R}_k^{(l+1)}
\end{equation}
where $\vec{R}^{(l)}_j$ represent the $R$-scores of the features of node $j$ at layer $l$, while the quantity $x_j A_{jk}$ models the extent to which node $j$ at layer $l$, with activation $x_j$, contributes to the relevance of node $k$ at layer $l+1$, where $A$ is the adjacency matrix. 


\section{Results}
\label{sec:Results}
\paragraph{Dataset}
\label{sec:Dataset}
We train the MLPF algorithm using a detector-agnostic \DELPHES~\cite{deFavereau:2013fsa} dataset from a previous publication~\cite{Pata:2021oez} found on the Zenodo platform~\cite{pata_joosep_2021_4559324} under CC BY 4.0 license.
It contains a particle-level training dataset of 50,000 top quark-antiquark (\ttbar) events produced in proton-proton collisions at 14\TeV generated with \PYTHIA8~\cite{Sjostrand:2006za,Sjostrand:2007gs} and \DELPHES3~\cite{deFavereau:2013fsa} from the HepSim software repository~\cite{Chekanov:2014fga}, overlaid with minimum bias events corresponding to 200 pileup collisions on average.
An additional testing dataset comprises 5,000 events composed uniquely of jets produced through the strong interaction, referred to as quantum chromodynamics (QCD) multijet events, with the same pileup conditions, is used for testing the model's performance and explainability.
Additional details can be found in Ref.~\cite{Pata:2021oez}.

We apply LRP to the fully trained MLPF model\footnote{Code can be found in Ref.~\cite{Farouk_code}}.
We run the LRP evaluation on the full testing dataset for about two days on one Nvidia A100 GPU. 
For each output neuron of each node, we obtain a tensor of the same dimensions as the input tensor of detector element features, which contains the $R$-scores of all input features in the event for that particular output neuron.
We refer to it as a relevancy-tensor ($R$-tensor).
Explicitly, the $R$-tensor has dimensions $[|Y|, n_y, |X|, n_x]$, where $n_y$ ($n_x$) is the number of PF candidate (detector element) features.
Given the dynamic graph construction, we note only the $k=16$ nearest neighbors identified in the latent space will have non-zero $R$-scores. 
To examine the general decision-making of the model, we randomly sample up to five $R$-tensors for each of the five PF-candidate node classes per event, sort the nearest neighbors by relevance score and average over many events to produce relevancy-maps ($R$-maps). 

\begin{figure}[H]
    \includegraphics[width=1\textwidth]{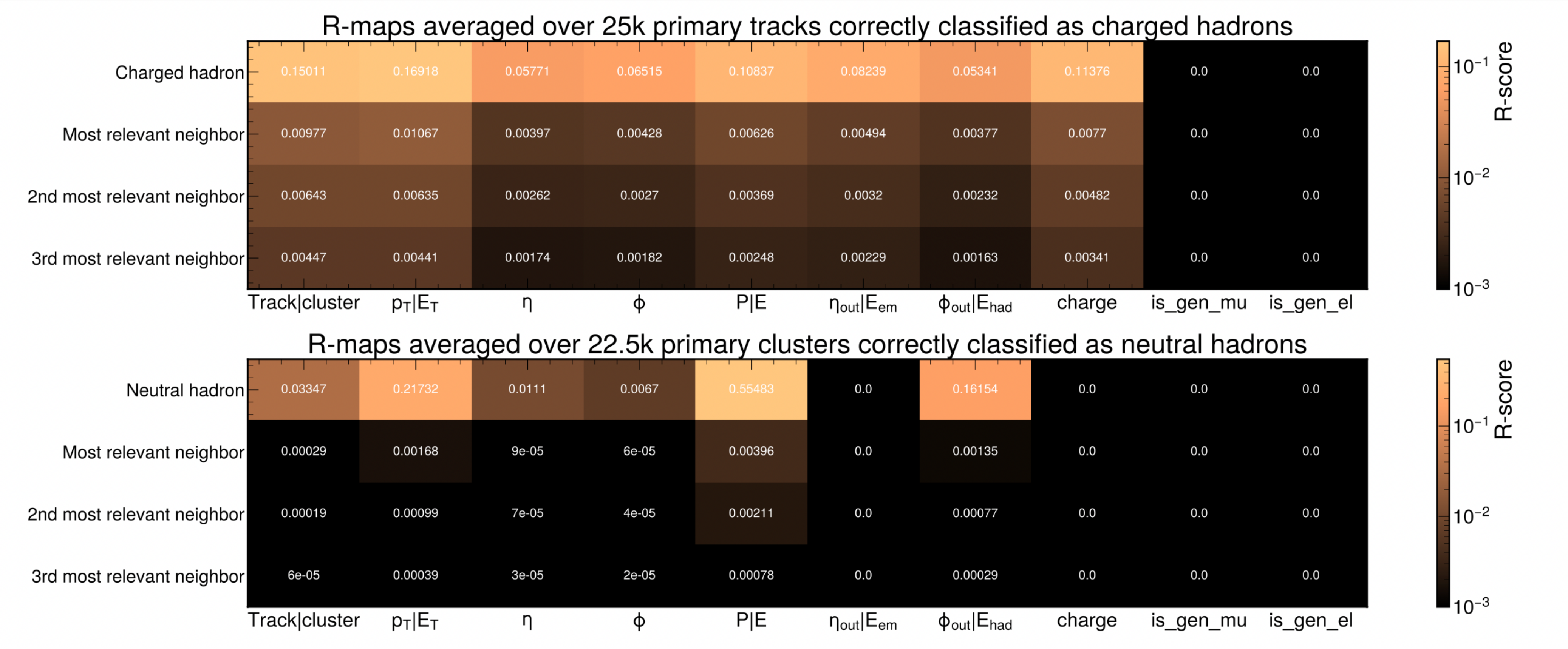}
    \caption{$R$-maps (with normalized $R$-scores computed using LRP) for elements associated to charged hadrons (top), and neutral hadrons (bottom). 
    The $R$-maps are those of the output neuron corresponding to the classification. 
    We see that charged hadrons use more neighbor information than neutral hadrons.}
    \label{fig:hadrons}
\end{figure}

\paragraph[R-maps]{$R$-maps}
Examples of $R$-maps for two common classes in our dataset, charged and neutral hadrons are presented in \autoref{fig:hadrons}.

\begin{figure}[H]
    \includegraphics[width=1\textwidth]{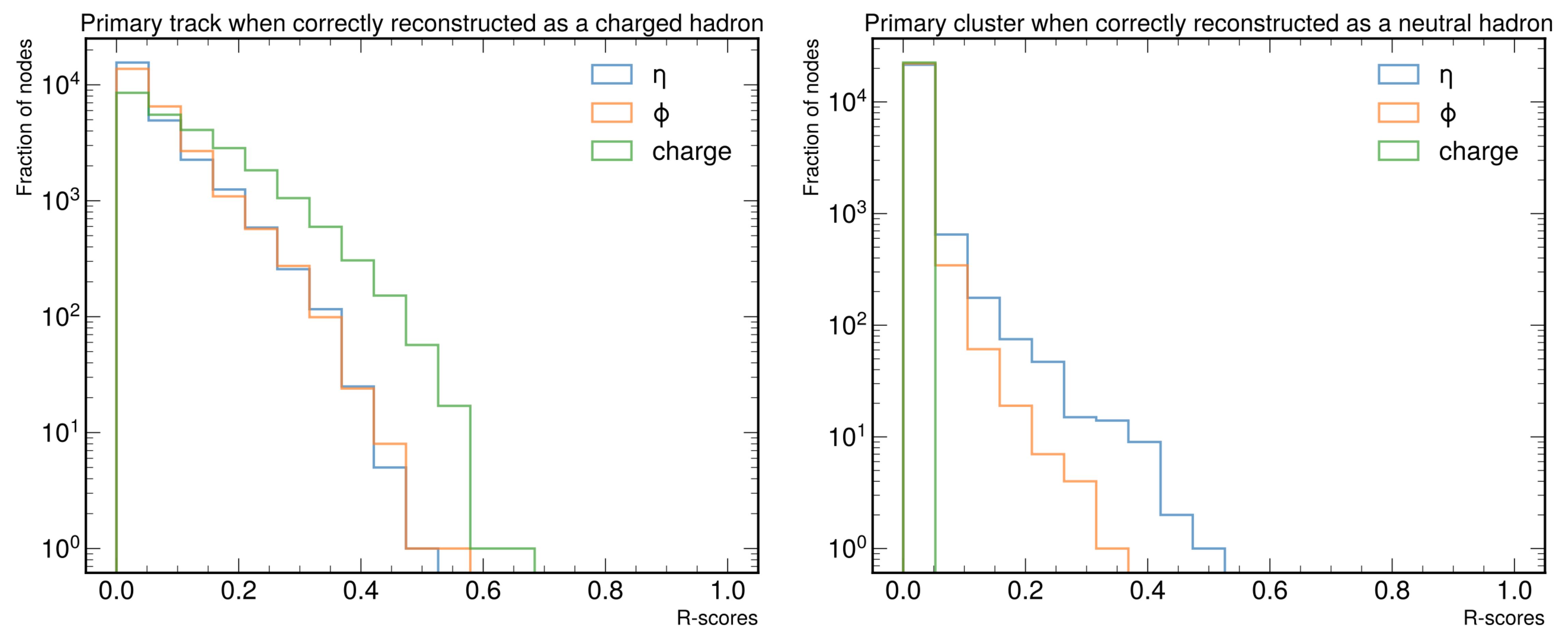}
    \caption{We focus on three input features: $\eta$, $\phi$, and charge, for clarity in the histograms.
    The charge feature is significantly more relevant for identifying a charged hadron compared to a neutral hadron.}
    \label{fig:hists}
\end{figure}

\paragraph{Histograms}
\label{sec:LRPhists}
Another way of visualizing the results of the $R$-maps is to create histograms for each output neuron for each of the classes. 
Examples of such histograms for charged and neutral hadrons are given in \autoref{fig:hists}.
Similar to the $R$-maps we have shown before, we only focus on the classification output neurons.
As expected, we find that the charge feature is significantly more relevant for identifying a charged hadron as compared to a neutral hadron. 
Although it is unsurprising that LRP reaches this conclusion, this work paves the way for exploring other, potentially less obvious, connections that can be surmised using LRP.

\section{Summary}
\label{sec:summary}
We implement the LRP technique on the MLPF algorithm, a GNN developed for the task of PF reconstruction. 
We find that, on average, correctly classified charged hadrons make significantly more use of neighboring information than correctly classified neutral hadrons. 
We suspect that this is because charged hadrons may give rise to multiple tracks (i.e. a primary associated track and its neighbors), which all need to be considered to correctly identify the charged hadron.
However, since our model only contains one message passing layer, it might be difficult, in general, for neighbors to have large relevant contributions.
This motivates us to further explore models with several message passing layers and check whether more neighbors will become relevant, for neutral hadron classification for instance, and how this would impact the model performance. 
It is also interesting to explore in more detail the distribution of the most relevant input features across the different classes to start obtaining physics or detector insights that might not be obvious from a traditional standpoint.
Finally, it would be interesting to explore LRP on the MLPF model when trained on a dataset more realistic  than \DELPHES, i.e, a full detector simulation including all material and physics effects.


\paragraph{Broader Impact}
\label{sec:impacts}
Physicists are usually wary of complex ML algorithms because the idea of a ``black box'' method may not be conducive to scientific discovery and understanding. 
This effort attempts to carefully examine how an ML model makes its decisions for a challenging particle physics task. 
This allows us to potentially better trust the ML model, and even learn from it. 
Furthermore, LRP can be potentially used as a pruning method to limit the resources required to perform PF-reconstruction. 
This would be very helpful in view of the planned high-luminosity upgrade of the CERN LHC.

\begin{ack}
F.~M. is supported by an Hal{\i}c{\i}o\u{g}lu Data Science Institute (HDSI) fellowship and an Institute for Research and Innovation in Software for High Energy Physics (IRIS-HEP) fellowship through the U.S. National Science Foundation (NSF) under Cooperative Agreement OAC-1836650.
F.~M., R.~K., D.~D., and J.~D. are supported by the US Department of Energy (DOE), Office of Science, Office of High Energy Physics Early Career Research program under Award No. DE-SC0021187 and by the DOE, Office of Advanced Scientific Computing Research under Award No. DE-SC0021396 (FAIR4HEP).
R.~K. is also supported by the LHC Physics Center at Fermi National Accelerator Laboratory, managed and operated by Fermi Research Alliance, LLC under Contract No. DE-AC02-07CH11359 with the DOE.
J.~P. is supported by the Mobilitas Pluss Grant No. MOBTP187 of the Estonian Research Council.
M.~P. is supported by the European Research Council (ERC) under the European Union's Horizon 2020 research and innovation program (Grant Agreement No. 772369).
J-R.~V. is supported by the DOE, Office of Science, Office of High Energy Physics under Award No. DE-SC0011925, DE-SC0019227, and DE-AC02-07CH11359.
J-R.~V. is additionally supported by the same ERC grant as M.~P.
This work was performed using the Pacific Research Platform Nautilus HyperCluster supported by NSF awards CNS-1730158, ACI-1540112, ACI-1541349, OAC-1826967, the University of California Office of the President, and the University of California San Diego's California Institute for Telecommunications and Information Technology/Qualcomm Institute. 
Thanks to CENIC for the 100\,Gpbs networks.
Funding for cloud credits was supported by NSF Award \#1904444 Internet2 Exploring Clouds to Accelerate Science (E-CAS).
\end{ack}

\bibliographystyle{lucas_unsrt}
\bibliography{bibliography}
\clearpage

\clearpage
\appendix

\section{LRP for electrons and muons}
\label{app:lrpleptons}
An interesting feature of the \DELPHES simulation is that it has access to generator level information that otherwise would not be accessible. 
For example, electrons and muons observed in the tracker will respectively have input features $\mathrm{is\_gen\_el}$ and $\mathrm{is\_gen\_mu}$. 
This information is necessary for the simple rule-based PF algorithm in \DELPHES, for example, to distinguish photons from electrons, due to the lack of full, realistic detector information.
Hence, with \DELPHES simulation, we normally restrict our view to the two classes: charged and neutral hadrons, when it comes to assessing the quality of reconstruction. 

Nevertheless, for this effort, it is interesting to look at the $R$-maps of electrons and muons as a way of verifying our LRP implementation.
\begin{figure}[H]
    \includegraphics[width=1\textwidth]{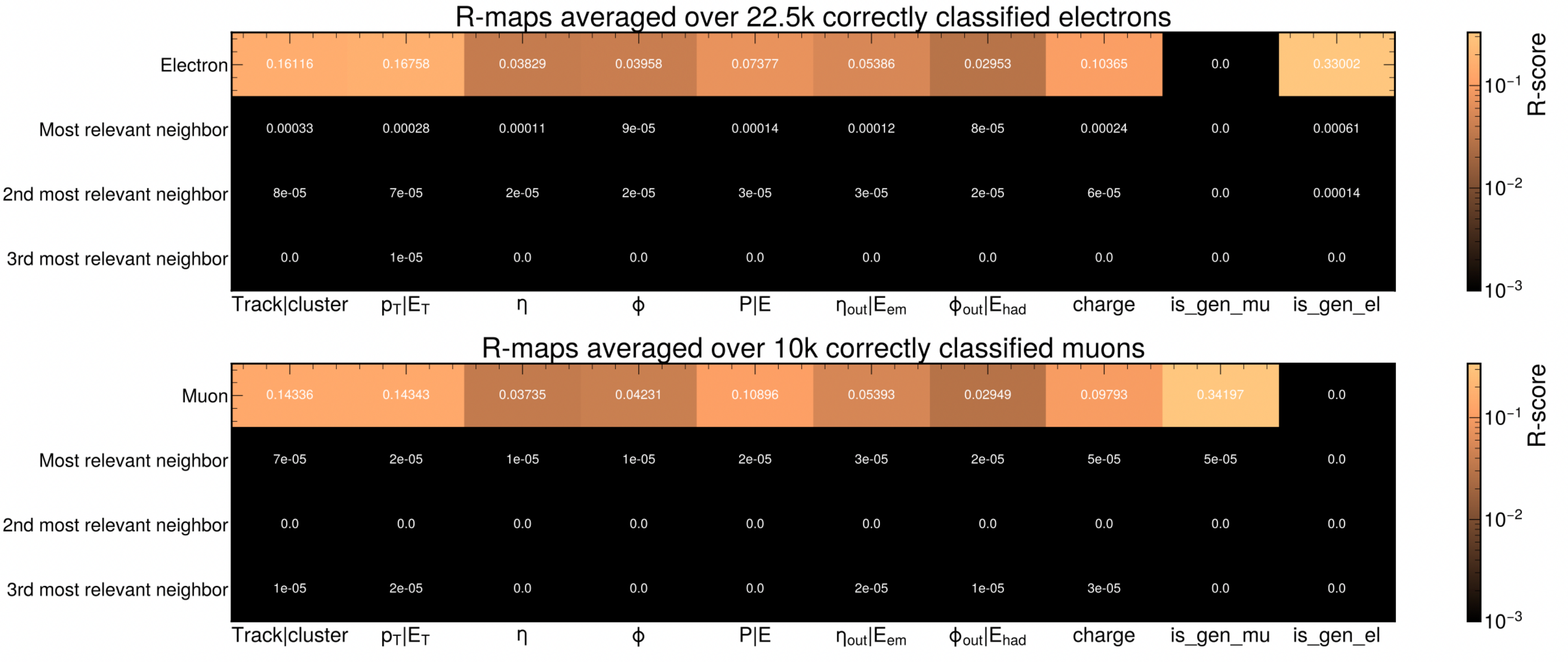}
    \caption{$R$-maps created for correctly classified electrons (top) and correctly classified muons (bottom). 
    We see that different features are highlighted for each different class, especially $\mathrm{is\_gen\_el}$ feature for correctly classified electrons, and $\mathrm{is\_gen\_mu}$ feature for correctly classified muons.}
    \label{fig:leptons}
\end{figure}
Looking at \autoref{fig:leptons}, we see that indeed LRP identifies $\mathrm{is\_gen\_el}$ and $\mathrm{is\_gen\_mu}$ for each of the electrons and muons respectively. 
Similar to the $R$-maps we have shown before, we only focus on the classification output neuron for either class.

\clearpage

\section{Model training and performance}
\label{app:performance}
For training the model, we use the same loss function as Ref.~\cite{Pata:2021oez}. 
We use the Adam algorithm with a learning rate of $10^{-4}$ for 300 epochs, training over $4\times10^{4}$ events, with $5\times10^{3}$ events used for testing. The events are processed in minibatches of four simultaneous events per graphics processing unit (GPU), we train for approximately 12 days using four RTX 1080Ti GPUs simultaneously. 
The modified architecture achieves comparable performance to the rule-based PF algorithm as shown in \autoref{app:performance}.

\begin{figure}[ht]
    \includegraphics[width=1\textwidth]{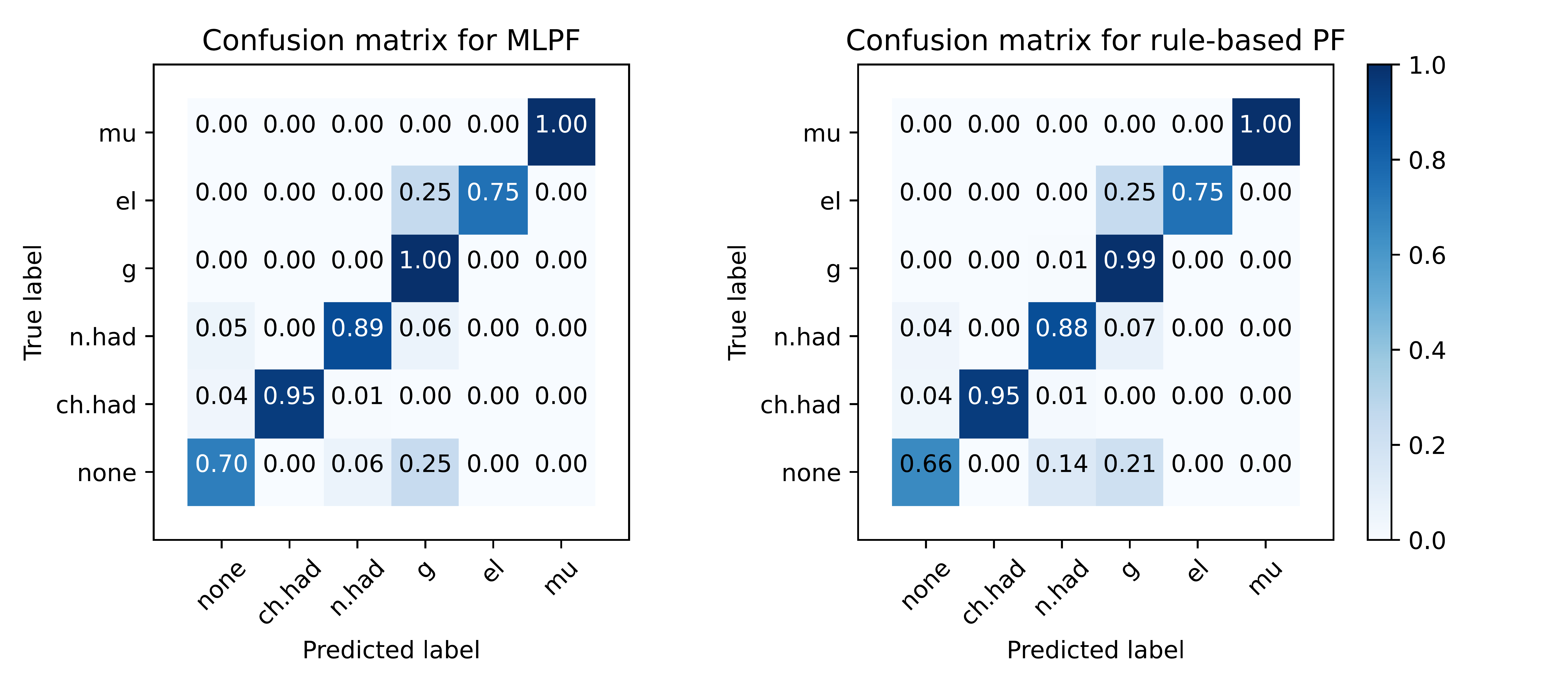}
    \caption{We observe similar classification performance when comparing the MLPF model's confusion matrix (left) to the rule-based PF algorithm's confusion matrix (right).}
    \label{fig:cm}
\end{figure}

\begin{figure}[ht]
    \includegraphics[width=1\textwidth]{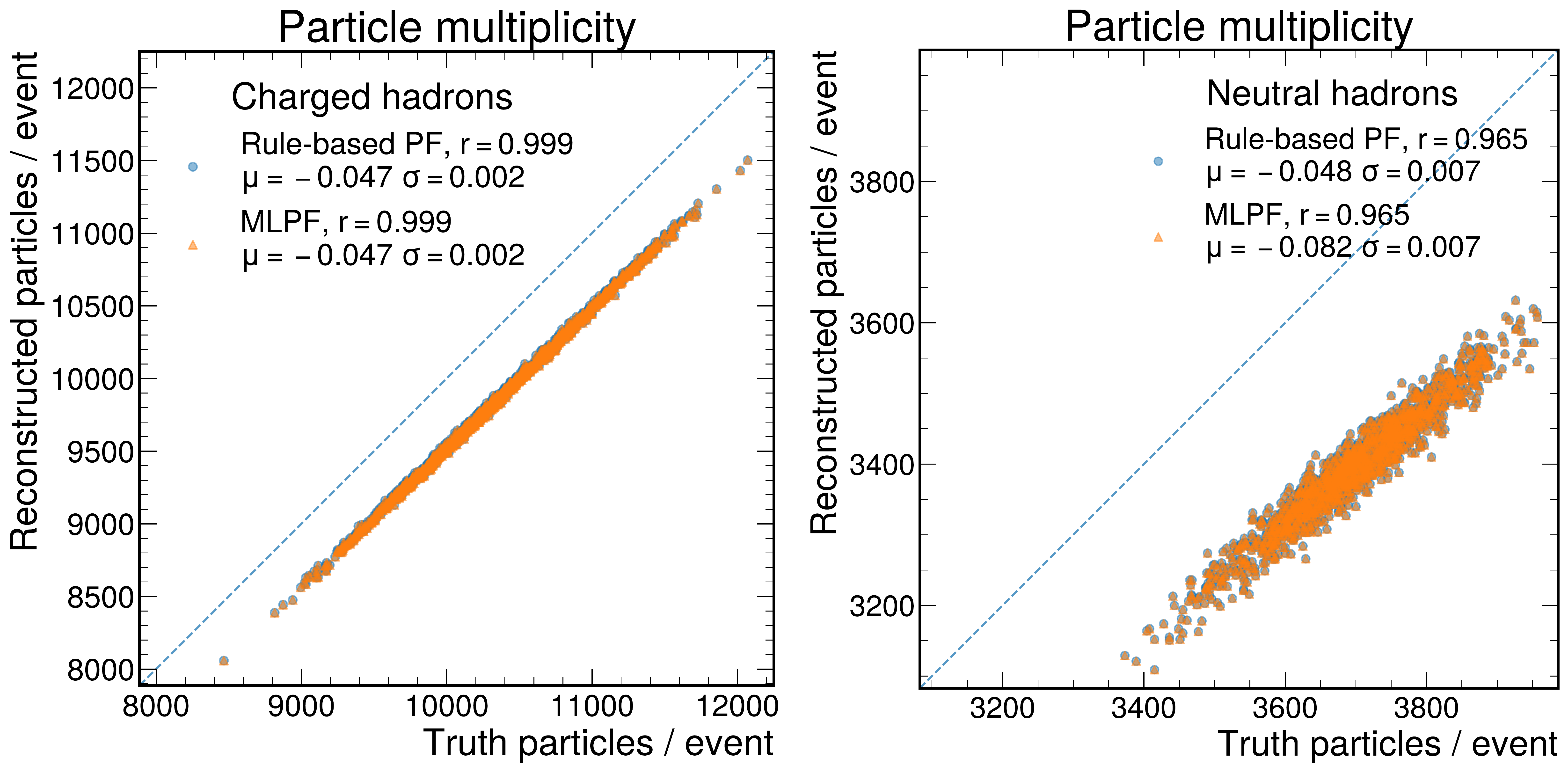}
    \caption{The MLPF model and rule-based PF algorithm perform similarly in predicting the particle multiplicity for charged (left) and neutral hadrons (right).}
    \label{fig:mult}
\end{figure}

The modified architecture of MLPF achieves comparable performance to the rule-based PF algorithm.
\autoref{fig:cm} shows the classification performance, quantified through a confusion matrix, is similar between the MLPF model and the rule-based PF algorithm.
Likewise, \autoref{fig:mult} shows similar ability of the two algorithms to predict the particle multiplicity for charged and neutral hadrons.

\end{document}